\documentclass[12pt]{article}
\usepackage{makeidx}
\usepackage{graphics}
\usepackage{amsthm}
\usepackage{amssymb}
\usepackage{amsmath}
\usepackage{amsfonts}
\usepackage{epic}
\usepackage{hyperref}
\usepackage{epsfig}
\usepackage{bm}
\usepackage{amscd}

\theoremstyle{plain}

\theoremstyle{definition}

\def\be{\begin{equation}}
\def\ee{\end{equation}}
\smallskip
\input{tcilatex}

\begin{document}

%\tableofcontents

\begin{titlepage}
\begin{flushright}
%hep-th/.......\\
\end{flushright}
%%%%%%%%%%%%%%%%%%%%%%%%%%%%%%%%%%%%%%%%%%%%%%%%%%%%%%%%%%%%%%%%%%%%%%%%
\begin{center}
\noindent{{\LARGE{On AGT description of ${\mathcal N}=2$ SCFT with $N_f=4$}}} 

\smallskip

\smallskip

\smallskip

\smallskip

\smallskip

\noindent{{\large{Gaston Giribet}}}

\smallskip
\smallskip

\smallskip

\smallskip
\smallskip
\noindent{{{Center for Cosmology and Particle Physics, New York University}}} 

\noindent{{{\it 4 Washington Place, NY10003, New York}}} 

\smallskip
\smallskip

\smallskip
\smallskip
\noindent{{{Consejo Nacional de Investigaciones Cient\'{\i}ficas y T\'{e}cnicas, CONICET}}} 

\noindent{{{\it Rivadavia 1917, 1033, Buenos Aires}}} 

\smallskip
\smallskip

\end{center}
\smallskip

\smallskip
\smallskip

\smallskip

\smallskip

\smallskip

\smallskip
\smallskip
\smallskip

\begin{abstract}
We consider Alday-Gaiotto-Tachikawa (AGT) realization of the Nekrasov partition function of ${\mathcal N}=2$ SCFT.
We focus our attention on the SU($2$) theory with $N_{f}= 4$ flavor symmetry, whose partition function, 
according to AGT, is given by the Liouville four-point function on the sphere. 
The gauge theory with $N_f = 4$ is known to exhibit SO($8$) symmetry. We explain how the Weyl symmetry transformations of
SO($8$) flavor symmetry are realized in the Liouville theory picture. This is associated to functional properties of the Liouville
four-point function that are {\it a priori} unexpected. In turn, this can be thought of as a non-trivial consistency check of AGT
conjecture. We also make some comments on elementary surface operators and WZW theory.
\end{abstract}

\end{titlepage}
%%%%%%%%%%%%%%%%%%%%%%%%%%%%%%%%%%%%%%%%%%%%%%%%%%%%%%%%%%%%%%%%%%%%%%%%

%\newpage

%\tableofcontents

%%%%%%%%%%%%%%%%%%%%%%%%%%%%%%%%%%%%%%%%%%%%%%%%%%%%%%%%%%%%%%%%%%%%%%

\section{Introduction}

\subsection{$\mathcal{N}=2$ theories}

In \cite{Gaiotto}, Gaiotto constructed a large family of SU($N_{c}$) $%
\mathcal{N}=2$ superconformal (quiver) gauge theories by compactifying the
(2,0) six-dimensional theory of the type A$_{N_{c}-1}$ on a two-dimensional
Riemann surface. This Riemann surface, which we denote by $\mathcal{C}_{g,n}$%
, is characterized by its genus $g$ and the weights of its $n$ punctures,
and in this way one gets a different $\mathcal{N}=2$ theory for each pair ($%
g,n$); we denote such gauge theory by $\mathcal{T}_{g,n}$.

Particular examples of the $\mathcal{N}=2$ theories that can be constructed
with this method are the $\mathcal{N}=4$ super Yang-Mills theory and the
so-called $\mathcal{N}=2^{\ast }$ theory, which correspond to compactifying
the six-dimensional theory on $\mathcal{C}_{1,0}$ and on $\mathcal{C}_{1,1}$
respectively. Another simple example is the compactification of the
six-dimensional theory on a 4-punctured sphere $\mathcal{C}_{0,4}$; in such
case one obtains the $\mathcal{N}=2$ gauge theory with $N_{f}=4$ flavor
symmetry, whose four mass parameters are given by the weights of the four
punctures.

In this construction, the modular parameters $\tau _{i}$ of $\mathcal{C}%
_{g,n}$ give the coupling constant $q_{i}$ of the corresponding quiver gauge
theory $\mathcal{T}_{g,n}$. In other words, the space of parameters of $%
\mathcal{T}_{g,n}$ is equivalent to the moduli space of complex structures
of $\mathcal{C}_{g,n}$. And it turns out that the group of duality
transformations of the gauge theory coincides with the mapping class group
of the Riemann surface $\mathcal{C}_{g,n}$. This is a very interesting
result as it permits to associate different ways of sewing the Riemann
surface $\mathcal{C}_{g,n}$ with different coupling limits of the gauge
theory $\mathcal{T}_{g,n}$. According to this picture, each way of sewing $%
\mathcal{C}_{g,n}$ as a set of $2g-2+n$ trinions connected by $3g-3+n$ tubes
is in correspondence with each Lagrangian description that the gauge theory $%
\mathcal{T}_{g,n}$ admits; in the limit where a given tube is thin and long,
the gauge theory description becomes weakly coupled.

\subsection{The AGT correspondence\ }

In a more recent paper \cite{AGT}, Alday, Gaiotto, and Tachikawa (AGT)\
reconsidered the $\mathcal{N}=2$ theories of \cite{Gaiotto} and studied the
Nekrasov partition function \cite{Nekrasov} associated to them. They arrived
to the following remarkable observation: In the case of those $\mathcal{N}=2$
theories whose gauge group corresponds to SU($2$), it happens that the
Nekrasov partition function on $\mathbb{R}^{4}$ is given by Virasoro
conformal blocks of Liouville field theory! This statement was confirmed by
explicit computation to several orders in a power expansion and the evidence
is quite convincing.

The result of \cite{AGT} seems to be revealing an intriguing connection
between four-dimensional superconformal gauge theories and two-dimensional
non-rational conformal theories. This was recently investigated in Refs. 
\cite%
{morozov2,morozov3,morozov4,morozov6,morozov8,bonelli,morozov9,morozov0,morozovA,otro1,Gaiotto2,fateev}%
, and extended in Refs. \cite%
{Vafa,Eguchi,Tachikawa,AldayTachikawa,morozov1,morozov5,morozov7,otro8,Schiappa}
to the case of SU($N_{c}$) gauge theories, for which the correspondence
seems to work with the A$_{N_{c}-1}$ Toda field theory. The results of \cite%
{AGT} also inspired very interesting works in the subject; see for instance
the recent \cite{Rastelli}. For early work on relations between $\mathcal{N}%
=2$ gauge theory and Liouville theory see \cite{Matone1} and references
therein.

The dictionary between the Nekrasov partition function of the $\mathcal{N}=2$
theory and the conformal blocks of Liouville theory is such that the two
deformation parameters $\varepsilon _{1,2}$ of the Nekrasov partition
function are given in terms of the Liouville central charge $c$ by the
relation $c=1+6Q^{2},$ with $Q=b+b^{-1}$ and $b=\varepsilon
_{1}=1/\varepsilon _{2}$. The external Liouville momenta $\alpha _{i}$ in
the $n$-point conformal blocks are given by mass parameters of the gauge
theory (see (\ref{mass3})-(\ref{mass2}) below). On the other hand, the
momenta of internal legs in the Liouville conformal blocks are given by the 
\textit{vev's} $a_{i}$ of the adjoint scalar fields in the $\mathcal{N}=2$
theory. See \cite{AGT} for details.

Even more surprising than the relation with Liouville conformal blocks is
the fact that the integral of the full Nekrasov partition function of a
given $\mathcal{N}=2$ theory $\mathcal{T}_{g,n}$ over $a_{i}$ turns out to
be given by the \textit{full} $n$-point correlation function of Liouville
field theory formulated on a genus-$g$ surface \cite{AGT}. Schematically, 
\begin{equation}
Z_{\text{Nekrasov}}\quad \underset{\text{AGT}}{\longleftrightarrow }\quad
\left\langle \prod\nolimits_{i=1}^{n}V_{\alpha _{i}}(z_{i})\right\rangle _{%
\text{Liouville}}.  \label{AGT}
\end{equation}

While in Liouville theory the correlation functions are defined by
assembling holomorphic and anti-holomorphic conformal blocks together and
integrating over the momenta of normalizable states in the intermediate
channels, in the gauge theory one integrates the modulus of the partition
function over the \textit{vev's} $a_{i}$ with an appropriate measure. (\ref%
{AGT}) somehow generalizes the Pestun's result for the theory on $\mathbb{S}%
^{4}$ with $b=1$ \cite{Pestum}.

One of the most interesting implications of the observations made in \cite%
{Gaiotto,AGT} is that the duality transformations in the $\mathcal{N}=2$
theories can be associated to geometric deformations of a Riemann surface,
in such a way that duality symmetry of the four-dimensional theory follows
from crossing symmetry of a two-dimensional CFT \cite{AGT}. It happens that
trinion decomposition of $\mathcal{C}_{g,n}$ codifies the connection between
different Lagrangian descriptions of $\mathcal{T}_{g,n}$ and their weakly
coupling regimes. Nevertheless, it turns out that not \textit{all} the
symmetries of a given $\mathcal{N}=2$ theory are necessarily associated to
crossing symmetry of Liouville correlators. In fact, as we will discuss
below, some symmetry transformations in the gauge theory turn out to be
captured by non-trivial functional properties of Liouville theory (which are
much less evident than crossing symmetry.)

\subsection{The theory with $N_{f}=4$}

In this paper, we will consider the case of SU($2$) $\mathcal{N}=2$ theory
with $N_{f}=4$ flavor symmetry. This gauge theory possesses global SO($8$)
symmetry. According to the dictionary of \cite{AGT}, the Nekrasov partition
function of this theory is given by the $4$-point function in Liouville
theory formulated on the sphere topology. \ The momenta $\alpha _{i}$ of the
four Liouville vertex operators $V_{\alpha _{i}}(z_{i})$ turn out to be
given by the mass parameters $m_{i}$ of the gauge theory by $\alpha
_{i}=Q/2+m_{i}$, and these parameters are related to the masses $\mu _{i}$
of the four hypermultiplets as follows%
\begin{eqnarray}
\mu _{1} &=&m_{1}+m_{2},\qquad \mu _{2}=m_{1}-m_{2},  \label{mass3} \\
\mu _{3} &=&m_{3}+m_{4},\qquad \mu _{4}=m_{3}-m_{4};  \label{mass2}
\end{eqnarray}%
which already corresponds to a particular trinion decomposition. On the
other hand, the gauge coupling constant of the gauge theory is given by the
cross-ratio $q=\frac{(z_{1}-z_{3})(z_{2}-z_{4})}{(z_{2}-z_{3})(z_{1}-z_{4})}$%
, where $z_{i}$ are the worldsheet locations of the Liouville vertex
operators.

The $\mathcal{N}=2$ theory with $N_{f}=4$ has SO($8$) global symmetry. The
masses $\mu _{i}$ of the hypermultiplets given in (\ref{mass3})-(\ref{mass2}%
) correspond to the diagonal Cartan elements of SO($8$), while $m_{i}$
correspond to the Cartan elements of its proper subgroup SO($4$)$\times $SO($%
4$). From the point of view of a given trinion decomposition, the
consideration of the subgroup SO($4$)$\times $SO($4$) is natural, as one has
SO($4$)$\simeq $SU($2$)$\times $SU($2$) corresponding to each pair of
hypermultiplets in each trinion.

One observes from (\ref{mass3})-(\ref{mass2}) that changing the sign of a
mass parameter like $m_{2}\rightarrow -m_{2}$ corresponds to exchanging the
masses of the hypermultiplets like $\mu _{1}\leftrightarrow \mu _{2}$. In
the Liouville theory picture this amounts to perform the reflection $\alpha
_{2}\rightarrow Q-\alpha _{2}$, which preserves the conformal dimension of
the corresponding vertex operator.

On the other hand, it turns out that the SL($2,\mathbb{Z}$) duality symmetry
of the gauge theory mixes with the SO($8$) triality in a funny way. The S$%
_{3}$ automorphism of Spin($8$) is generated by two transformations that act
on the mass $\mu _{i}$ in a non-diagonal way \cite{SeibergWitten}. On of
these transformations is

\begin{eqnarray}
\mu _{i} &\rightarrow &\mu _{i}-\frac{\mu _{1}+\mu _{2}+\mu _{3}-\mu _{4}}{2}%
,\qquad \text{for\ }i=1,2,3,  \label{cc1} \\
\mu _{4} &\rightarrow &\frac{\mu _{1}+\mu _{2}+\mu _{3}+\mu _{4}}{2},
\label{cc4}
\end{eqnarray}%
and we also have the transformation 
\begin{eqnarray}
\mu _{i} &\rightarrow &\mu _{i},\qquad \text{for\ }i=1,2,3,  \label{c6} \\
\mu _{4} &\rightarrow &-\mu _{4}.  \label{c7}
\end{eqnarray}%
While (\ref{c6})-(\ref{c7}) corresponds to the operation of exchanging the
two spinor representation of the group, transformation (\ref{cc1})-(\ref{cc4}%
) corresponds to the operation of exchanging the vector representation with
one of the spinor representations (physically, this corresponds to exchange
elementary particles and monopoles in the gauge theory.) These triality
transformations have a simple interpretation in terms of the Liouville
picture: From (\ref{mass3})-(\ref{mass2}) we observe that performing (\ref%
{cc1})-(\ref{cc4}) corresponds to exchange $m_{1}\leftrightarrow -m_{4}$
keeping $m_{2}$ and $m_{3}$ fixed. Since this amounts to exchange the
trinion decomposition, this means that triality is in correspondence with
crossing symmetry in the Liouville theory. Even simpler is the form that
transformation (\ref{c6})-(\ref{c7}) adopts in the Liouville side; it
corresponds to interchange $m_{3}\leftrightarrow m_{4}$ keeping $m_{1}$ and $%
m_{2}$ fixed.

From this we see that some simple transformations of SO($8$) can be
interpreted in a very natural way in the Liouville theory picture. However,
there are some special Weyl transformations whose identification as symmetry
in the Liouville theory side is much less evident, and this is what we want
to study here. One such a symmetry transformation is given by%
\begin{equation}
(\mu _{1},\mu _{2},\mu _{3},\mu _{4})\rightarrow (-\mu _{4},\mu _{2},\mu
_{3},-\mu _{1}),
\end{equation}%
which in terms of the parameters $m_{i}$ reads%
\begin{eqnarray}
m_{1} &\rightarrow &\frac{m_{1}-m_{2}-m_{3}+m_{4}}{2},\qquad
m_{2}\rightarrow \frac{-m_{1}+m_{2}-m_{3}+m_{4}}{2},  \label{c1} \\
m_{3} &\rightarrow &\frac{-m_{1}-m_{2}+m_{3}+m_{4}}{2},\qquad
m_{4}\rightarrow \frac{m_{1}+m_{2}+m_{3}+m_{4}}{2}.  \label{c4}
\end{eqnarray}%
This symmetry is not at all evident in the Liouville theory picture. In
fact, it corresponds to a non-diagonal transformation of the momenta $\alpha
_{i}$ (see (\ref{a2})-(\ref{a4}) below). In this paper we address the
question about how the symmetry under (\ref{c1})-(\ref{c4}) can be proven
from the Liouville point of view. We will show this symmetry is actually
realized in the Liouville theory side by non-trivial functional relations
between different four-point correlation functions. These relations are 
\textit{a priori} unexpected, and this is why this calculation is an
interesting consistency check of the proposal in \cite{AGT}.

\subsection{Overview}

The paper is organized as follows: In Section 2, we study the analytic
extension of the integral representation of Liouville correlation functions.
We use this representation to prove the symmetry under transformations (\ref%
{c1})-(\ref{c4}) from the Liouville point of view. In the 2D CFT language,
this symmetry is expressed by Eq. (\ref{Postalin}). Because we derive the
symmetry relation (\ref{Postalin}) by resorting exclusively to elements of
Liouville field theory, we like to think of it as a nice consistency check
of AGT conjecture. The full SO($8$) is not manifest in the six-dimensional
picture, but it emerges in the infrared. Therefore, the fact of proving the
Weyl symmetry (\ref{c1})-(\ref{c4}) in the Liouville picture is quite
interesting. In Section 3, we rederive (\ref{Postalin}) in an alternative
(less direct) way. The strategy goes as follows: First we consider a
five-point function in Liouville theory, which is meant to describe surface
operator in the $\mathcal{N}=2$ gauge theory. Then, we write this Liouville
five-point function as a four-point function in the Wess-Zumino-Witten
theory (WZW) with affine $\widehat{A}_{1}$ symmetry. Using functional
relations between different solutions to the Knizhnik-Zamolodchikov equation
and performing Hamiltonian reduction from WZW to Liouville we reobtain the
right symmetry relation with the appropriate coefficient. We also make some
comments on how to describe the surface operator of the $\mathcal{N}=2$
theory in terms of the $\widehat{A}_{1}$ WZW theory.

\section{Liouville theory and Weyl symmetry}

\subsection{Integral representation of correlation functions}

Liouville theory is defined by the action \cite{Seiberg,Yu,TeschnerReview}%
\begin{equation}
S_{L}[\varphi ]=\frac{1}{4\pi }\int d^{2}z\left( \partial \varphi \overline{%
\partial }\varphi +\frac{1}{2\sqrt{2}}QR\varphi +4\pi \mu e^{\sqrt{2}%
b\varphi }\right)  \label{mancha}
\end{equation}%
where $\mu $ is a positive constant. The background charge $Q$ takes the
value $Q=b+b^{-1}$ for the Liouville self-potential $\mu e^{\sqrt{2}b\varphi
}$ to be a marginal operator. The theory is globally defined after one
specifies the boundary conditions. For the theory on the sphere, one imposes
the asymptotic behavior $\varphi \sim -2\sqrt{2}Q\log |z|$ for large $|z|$.
In the conformal gauge, the linear dilaton term $QR\varphi $ is understood
as receiving a contribution from the point at infinity due to the scalar
curvature $R$ of the sphere.

Under holomorphic transformations $z\rightarrow w$ Liouville field $\varphi $
transforms like $\varphi \rightarrow \varphi -\sqrt{2}Q\log |\frac{dw}{dz}|$%
. The central charge receives a contribution from the background charge, and
it is%
\begin{equation*}
c=1+6Q^{2}.
\end{equation*}

Here we are interested in the exponential vertex operators of the theory 
\cite{TeschnerExponencial}%
\begin{equation}
V_{\alpha }(z)=e^{\sqrt{2}\alpha \varphi (z)}.  \label{V}
\end{equation}%
These are local operators that create primary states of conformal dimension $%
\Delta _{\alpha }=\alpha (Q-\alpha )$. Notice that $\Delta _{\alpha }$
remains invariant under $\alpha \rightarrow \alpha ^{\ast }=Q-\alpha $,
which means that vertices $V_{\alpha }(z)=e^{\sqrt{2}\alpha \varphi (z)}$
and $V_{\alpha ^{\ast }}(z)=e^{\sqrt{2}(Q-\alpha )\varphi (z)}$ have the
same conformal dimension.

Correlation functions in Liouville field theory are defined by the
expectation value of a product of vertex operators (\ref{V}), namely%
\begin{equation*}
\Omega ^{(n)}(\alpha _{1},...\alpha _{n}|z_{1},...z_{n})\ \equiv \
\left\langle \prod\nolimits_{i=1}^{n}V_{\alpha _{i}}(z_{i})\right\rangle _{%
\text{Liouville}}=\int \mathcal{D}\varphi \ e^{-S_{L}[\varphi
]}\prod_{i=1}^{n}e^{\sqrt{2}\alpha _{i}\varphi (z_{i})}.
\end{equation*}

Integrating over the zero mode of the field, one finds the following
expression \cite{GL,DornOtto} 
\begin{equation}
\Omega ^{(n)}(\alpha _{1},...\alpha _{n}|z_{1},...z_{n})=\frac{\Gamma (-s)}{b%
}\mu ^{s}\int \prod_{r=1}^{s}d^{2}w_{r}\ \Omega ^{(n+s)}(\alpha
_{1},...\alpha _{n},b,...b|z_{1},...z_{n},w_{1},...w_{s})_{|\mu =0},
\label{U}
\end{equation}%
where $s=-b^{-1}(\alpha _{1}+\alpha _{2}+...\alpha _{n})+1+b^{-2}$, and
where the average on the right hand side is defined in terms of the free
theory $\mu =0$. The factor $\Gamma (-s)$ in (\ref{U}) arises from the
integration over the zero-mode of $\varphi $ \cite{GL,DiFrancescoKutasov},
and this also gives \ a $\delta $-function that completely determines the
amount of screening operators $V_{b}(w)$ that appear in the non-vanishing
correlators. In deriving (\ref{U}), the Gauss-Bonnet theorem is used to
determine the relation between $s$, $b$, and the momenta $\alpha _{i}$. For
a genus-$g$ $n$-puncture Riemann surface the relation is%
\begin{equation}
bs=Q(1-g)-\sum_{i=1}^{n}\alpha _{i}.
\end{equation}

Then, (\ref{U}) permits to compute correlators by performing the Wick
contraction of the $n+s$ operators and using the free field propagator $%
\left\langle \varphi (z_{1})\varphi (z_{2})\right\rangle =-2\log
|z_{1}-z_{2}|$. This yields%
\begin{equation}
\Omega ^{(n)}(\alpha _{1},...\alpha _{n}|z_{1},...z_{n})=\Gamma (-s)\Gamma
(s+1)b^{-1}\mu ^{s}\ \mathcal{I}_{s}^{(n)}(\alpha _{1},\alpha _{2},...\alpha
_{n}|z_{1},z_{2},...z_{n}),  \label{n9}
\end{equation}%
with%
\begin{equation}
\ \mathcal{I}_{s}^{(n)}(\alpha _{1},\alpha _{2},...\alpha
_{n}|z_{1},z_{2},...z_{n})=\frac{1}{s!}\prod_{i<j}^{n}|z_{i}-z_{j}|^{-4%
\alpha _{i}\alpha _{j}}\int
\prod_{r=1}^{s}d^{2}w_{r}\prod_{i=1}^{n}\prod_{r=1}^{s}|z_{i}-w_{r}|^{-4b%
\alpha _{i}}\prod_{r<t}^{s}|w_{t}-w_{r}|^{-4b^{2}},  \label{DF}
\end{equation}%
where each integral is over the whole complex plane $\mathbb{C}$.

Expression (\ref{DF}) has to be considered just formally. This is because,
in general, the amount of screening operators $s$ is not an integer number.
Therefore, in order to compute generic correlation functions one has to deal
first with the problem of making sense of the integral representation (\ref%
{n9}). In the case of $n$-point functions with $n\leq 3$ and $s\in \mathbb{Z}%
_{>0}$, multiple integral (\ref{DF}) can be solved using the results of
Refs. \cite{DF,DF2}. Generic Liouville correlation functions are thus
defined by analytic extension. This analytic extension is accomplished by
continuing the multiple integral (\ref{DF}) to non-integer (and non-real)
values of $s$. The extension to non-integer $s$ was discussed in the
literature, and it plays a crucial r\^{o}le in our discussion. To see how it
works, let us consider the calculation of the Liouville partition function
on the sphere as an illustrative example: This corresponds to $g=0$ and $n=0$%
. The number of screening operators in this case is $m=s-3=-2+b^{2}$, and
this is because, in order to compute the genus-zero zero-point function, one
has to consider the correlator with three local operators $e^{\sqrt{2}%
b\varphi (z)}$ inserted at fixed points $z_{1}=0,z_{2}=1$, and $z_{3}=\infty 
$; this amounts to compensate the volume of the conformal Killing group. In
turn, genus-zero Liouville partition function is given by 
\begin{equation*}
\Omega ^{(0)}=\ \mu ^{m+3}b^{-1}\Gamma (-m-3)\Gamma (m+1)\ \mathcal{I}%
_{m}^{(3)}(b,b,b|0,1,\infty ),
\end{equation*}%
with $m=-2+b^{-2}$. If $m$ was a positive integer number (what happens only
if $b^{2}\in \mathbb{Z}_{>2}$) this integral could be solved by using the
Dotsenko-Fateev integral formulas of \cite{DF}. However, here we are
interested in the case where $m$ is generic enough. The way one circumvents
this obstruction is assuming the condition $m\in \mathbb{Z}_{>0}$ through
the integration and then analytically extending the final expression. More
precisely, one integrates $\mathcal{I}_{m}^{(3)}(b,b,b|0,1,\infty )$ and
obtains%
\begin{equation*}
\Omega ^{(0)}=\frac{\mu ^{3+m}}{b}\Gamma (-m-3)\Gamma (m+1)\pi ^{m}\gamma
^{m}(1+b^{2})\prod_{r=1}^{m}\gamma (-rb^{2})\prod_{r=0}^{m-1}\gamma
^{2}(1-(2+r)b^{2})\gamma (-1+(3+r+m)b^{2}).
\end{equation*}%
where $\gamma (x)=\Gamma (x)/\Gamma (1-x)$, which, again, is an expression
that seems to make sense only if $m\in \mathbb{Z}_{>0}$. Then, one extends
the result to the whole range $b^{2}\in \mathbb{R}_{>1}$ by rewriting the
products in the expression above appropriately. This is done by noticing
that $\gamma (-1+(3+r+m)b^{2})=\gamma ((r+1)b^{2})$, and using that $%
m=-2+b^{-2}$ and $1-rb^{2}=(m+2-r)b^{2}$. Rearranging the product of $\gamma 
$-functions and using properties of the $\Gamma $-function, one finally finds%
\begin{equation}
\Omega ^{(0)}=\frac{(1-b^{2})\left( \pi \mu \gamma (b^{2})\right) ^{Q/b}}{%
\pi ^{3}Q\gamma (b^{2})\gamma (b^{-2})},  \label{0pf}
\end{equation}%
which is the exact result for the Liouville partition function on the sphere.

The two-point correlation functions can be computed in a similar way,
analytically continuing the integral formula for $\mathcal{I}%
_{b^{-2}-2\alpha b^{-1}}^{(3)}(\alpha ,\alpha ,b|0,1,\infty )$. The result
reads%
\begin{equation}
\Omega ^{(2)}(\alpha ,\alpha |0,1)=\left( \pi \mu \gamma (b^{2})\right)
^{(Q-2\alpha )/b}\frac{\gamma (2\alpha b-b^{2})\gamma (2\alpha b^{-1}-b^{-2})%
}{\pi (2\alpha -Q)},\quad \quad \Omega ^{(2)}(\alpha ,\alpha ^{\ast }|0,1)=1.
\label{2pf}
\end{equation}

Up to a factor $(Q-2\alpha )/\pi $, the two-point function $\Omega
^{(2)}(\alpha ,\alpha |0,1)$ coincides with the Liouville reflection
coefficient, which yields the functional relation 
\begin{equation}
\Omega ^{(n)}(\alpha _{1},...\alpha _{n}|z_{1},...z_{n})=\frac{\pi }{%
Q-2\alpha _{n}}\ \Omega ^{(2)}(\alpha _{n},\alpha _{n}|0,1)\ \Omega
^{(n)}(\alpha _{1},...,\alpha _{n-1},\alpha _{n}^{\ast }|z_{1},...z_{n})
\label{reflection}
\end{equation}

As zero-point and two-point functions, the three-point function can also be
computed by performing the appropriate analytic extension of the product of $%
\Gamma $-functions in the Dotsenko-Fateev integral formula. To do this
properly, one has first to be reminded of the fact that Liouville theory
exhibits self-duality under $b\leftrightarrow b^{-1}$ and that this
self-duality is associated to the existence of a second screening operator $%
V_{1/b}(w)=e^{\sqrt{2}b^{-1}\varphi (w)}$ in the theory. Taking this into
account, one has enough information to reconstruct the \textit{full} pole
structure of the exact three-point function\ and then analytically extend
integral (\ref{DF}). This amounts to introduce the $\Upsilon $-function \cite%
{ZZ} 
\begin{equation}
\log \Upsilon _{b}(x)=\int_{\mathbb{R}_{>0}}\frac{dt}{t}\left( \left( \frac{Q%
}{2}-x\right) ^{2}e^{-t}-\frac{\sinh ^{2}\left( \left( \frac{Q}{2}-x\right) 
\frac{t}{2}\right) }{\sinh \left( \frac{tb}{2}\right) \sinh \left( \frac{t}{%
2b}\right) }\right) ,  \label{Upsilon}
\end{equation}%
which presents poles at $x=mb+nb^{-1}$ and $x=-(m+1)b-(n+1)b^{-1}$ with $%
m,n\in \mathbb{Z}_{>0}$, which manifestly shows the symmetry of the pole
structure under $b\leftrightarrow b^{-1}$. This function obeys the self-dual
properties 
\begin{equation}
\Upsilon _{b}(x)=\Upsilon _{1/b}(x),\quad \quad \quad \Upsilon
_{b}(x)=\Upsilon _{b}(Q-x),  \label{ref}
\end{equation}%
and the shift properties%
\begin{equation}
\Upsilon _{b}(x+b)=b^{1-2bx}\gamma (bx)\Upsilon _{b}(x),\quad \quad \Upsilon
_{b}(x+b^{-1})=b^{-1+2x/b}\gamma (x/b)\Upsilon _{b}(x).  \label{shift}
\end{equation}%
In particular, if we define the function $P(m)=\prod_{r=1}^{m}\gamma
(rb^{2}) $ for $m\mathbb{\in Z}_{>0},$ and $P(0)=1,$ the shift properties
imply that 
\begin{equation}
P(m)=\frac{\Upsilon _{b}(mb+b)}{\Upsilon _{b}(b)}b^{m((m+1)b^{2}-1)},\qquad
m\in \mathbb{Z}_{\geq 0}.  \label{Lapo}
\end{equation}%
This admits the following extension to negative values of $m$%
\begin{equation}
P(m)=b^{-4(m+1)}\frac{\gamma (-m)}{\gamma (-mb^{2})}P(-m),\qquad m\in 
\mathbb{Z}_{<0};  \label{dddale}
\end{equation}%
see \cite{Lorena} and references therein.

Then, writing the products of $\Gamma $-functions that arise in the
Dotsenko-Fateev integral formula for $n=3$ in terms of $\Upsilon $-functions
using (\ref{Lapo}), one finally obtains the exact three-point function \cite%
{DornOtto2,ZZ,Teschner3pf,Pakman}; namely%
\begin{equation}
\Omega ^{(3)}(\alpha _{1},\alpha _{2},\alpha _{3}|0,1,\infty )=\left( \pi
\mu \gamma (b^{2})b^{2-2b^{2}}\right) ^{(Q-\alpha )/b}\frac{\Upsilon _{b}(b)%
}{\Upsilon _{b}(\alpha -Q)}\prod_{i=1}^{3}\frac{\Upsilon _{b}(2\alpha _{i})}{%
\Upsilon _{b}(\alpha -2\alpha _{i})},  \label{3pf}
\end{equation}%
where $\alpha =\alpha _{1}+\alpha _{2}+\alpha _{3}$.

The computation of (\ref{0pf}), (\ref{2pf}) and (\ref{3pf}) shows that the
method of analytically continuing the integral representation of correlation
functions works and it is a powerful tool. We will use this method to derive
(\ref{Postalin}) below.

\subsection{Four-point function and Weyl symmetry}

In this paper, and because we are interested in the AGT description of the $%
\mathcal{N}=2$ gauge theory with $N_{f}=4$, we are interested in the
Liouville four-point function. Four-point function is substantially more
complicated than the cases $n=0,2,3$. In that case, the integral
representation takes the form 
\begin{equation}
\Omega ^{(4)}(\alpha _{1},\alpha _{2},\alpha _{3},\alpha _{4}|0,1,\infty
,q)=\Gamma (-s)\Gamma (s+1)b^{-1}\mu ^{s}\ \mathcal{I}_{s}^{(4)}(\alpha
_{1},\alpha _{2},\alpha _{3},\alpha _{4}|q),
\end{equation}%
with $s=-b^{-1}(\alpha _{1}+\alpha _{2}+\alpha _{3}+\alpha _{4}-Q)$ and 
\begin{eqnarray}
\mathcal{I}_{s}^{(4)}(\alpha _{1},\alpha _{2},\alpha _{3},\alpha _{4}|q) &=&%
\frac{|q|^{-4\alpha _{1}\alpha _{4}}|1-q|^{-4\alpha _{2}\alpha _{4}}}{\Gamma
(s+1)}\int \prod_{r=1}^{s}d^{2}w_{r}\prod_{r=1}^{s}|w_{r}|^{-4b\alpha
_{1}}|1-w_{r}|^{-4b\alpha _{2}}|w_{r}-q|^{-4b\alpha _{4}}\times  \notag \\
&&\prod_{r<t}^{s}|w_{t}-w_{r}|^{-4b^{2}},  \label{n4}
\end{eqnarray}%
with $z_{1}=0,$ $z_{2}=1$, $z_{3}=\infty ,$ and $z_{4}=q$. It was shown in 
\cite{FL2} that, if $2\alpha _{4}/b=-m\in \mathbb{Z}_{<0}$, then the
integral (\ref{n4}) satisfies the following remarkable property%
\begin{eqnarray}
\mathcal{I}_{s}^{(4)}(\alpha _{1},\alpha _{2},\alpha _{3},\alpha _{4}|q)
&=&|q|^{4(\widetilde{\alpha }_{1}\widetilde{\alpha }_{4}-\alpha _{1}\alpha
_{4})}|1-q|^{4(\widetilde{\alpha }_{2}\widetilde{\alpha }_{4}-\alpha
_{2}\alpha _{4})}(-\pi \gamma (1+b))^{s-m}\prod_{r=1}^{s-m}\gamma (2b\alpha
_{4}-rb^{2})\times  \notag \\
&&\prod_{r=0}^{s-m-1}\prod_{i=1}^{3}\gamma (1-2b\alpha _{i}-rb^{2})\ 
\mathcal{I}_{m}^{(4)}(\widetilde{\alpha }_{1},\widetilde{\alpha }_{2},%
\widetilde{\alpha }_{3},\widetilde{\alpha }_{4}|q),  \label{IntegralRelation}
\end{eqnarray}%
with 
\begin{eqnarray}
\widetilde{\alpha }_{1} &=&\frac{Q}{2}+\frac{\alpha _{1}-\alpha _{2}-\alpha
_{3}+\alpha _{4}}{2},\qquad \widetilde{\alpha }_{2}=\frac{Q}{2}-\frac{\alpha
_{1}-\alpha _{2}+\alpha _{3}-\alpha _{4}}{2},  \label{a2} \\
\widetilde{\alpha }_{3} &=&\frac{Q}{2}-\frac{\alpha _{1}+\alpha _{2}-\alpha
_{3}-\alpha _{4}}{2},\qquad \widetilde{\alpha }_{4}=-\frac{Q}{2}+\frac{%
\alpha _{1}+\alpha _{2}+\alpha _{3}+\alpha _{4}}{2}.  \label{a4}
\end{eqnarray}

The proof of (\ref{IntegralRelation}), for $m\in \mathbb{Z}_{>0}$, follows
from iterating recursion relations for $\mathcal{I}_{s}^{(n)}(\alpha
_{1},...\alpha _{n}|z_{1,}...z_{n})$; see \cite{FL1,FL2} for details.

Expressions (\ref{a2})-(\ref{a4}) anticipate the point we want to make here:
Since the mass of the hypermultiplets in the gauge theory $\mu _{i}$ are
given by $\alpha _{i}=\frac{Q}{2}+m_{i}$, provided with (\ref{cc1})-(\ref%
{cc4}), then equation (\ref{IntegralRelation}) seems to incarnate some kind
of invariance under (\ref{c1})-(\ref{c4}) that Liouville four-point function
exhibits. To make it precise, what we have to do first is to analytically
continue the integral relation (\ref{IntegralRelation}) to complex values of 
$\alpha _{4}$, and then give one such a relation for the \textit{exact}
four-point function. The analytic continuation is accomplished by following
a recipe: First, we may use expression (\ref{Lapo}); then, we have to
remember that a product $\prod_{r=1}^{m}F(r)$ can be extended to the range $%
m\in \mathbb{Z}_{<0}$ by replacing it by the expression by $%
\prod_{r=0}^{-m-1}F^{-1}(-r)$, which, in particular, amounts to give the
formula (\ref{dddale}) for $P(m)$ with $m\in \mathbb{Z}_{<0}$. This,
together with the convenient use of properties of $\Gamma $-functions, leads
us to the following remarkable equation%
\begin{equation}
|q|^{4\widetilde{\alpha }_{1}\widetilde{\alpha }_{4}}|1-q|^{4\widetilde{%
\alpha }_{2}\widetilde{\alpha }_{4}}\ \frac{\Omega ^{(4)}(\widetilde{\alpha }%
_{1},\widetilde{\alpha }_{2},\widetilde{\alpha }_{3},\widetilde{\alpha }%
_{4}|0,1,\infty ,q)}{f(\widetilde{\alpha }_{1})f(\widetilde{\alpha }_{2})f(%
\widetilde{\alpha }_{3})f(\widetilde{\alpha }_{4}^{\ast })}=|q|^{4\alpha
_{1}\alpha _{4}}|1-q|^{4\alpha _{2}\alpha _{4}}\ \frac{\Omega ^{(4)}(\alpha
_{1},\alpha _{2},\alpha _{3},\alpha _{4}|0,1,\infty ,q)}{f(\alpha
_{1})f(\alpha _{2})f(\alpha _{3})f(\alpha _{4}^{\ast })}  \label{Postalin}
\end{equation}%
with%
\begin{equation*}
f(\alpha _{i})=\left( \pi \mu \gamma (b^{2})b^{2-2b^{2}}\right) ^{-\alpha
_{i}/2b}\Upsilon _{b}(2\alpha _{i}).
\end{equation*}%
Notice that $\Upsilon _{b}(2\alpha _{i})=\Upsilon _{b}(2\alpha _{i}^{\ast })$%
. To prove (\ref{Postalin}) we also used the functional properties of the $%
\Upsilon $-function (\ref{ref})-(\ref{shift}) and the fact that its
derivative obeys $\Upsilon _{b}^{\prime }(-mb)=(-1)^{m}\Gamma (m+1)\Gamma
(-m)b^{-1}\Upsilon _{b}(-mb)$. It is important to distinguish between
identity (\ref{IntegralRelation}) and (\ref{Postalin}). (\ref{Postalin}) is
meant to hold between \textit{exact} four-point functions, and, unlike (\ref%
{IntegralRelation}), is manifestly symmetric under $\widetilde{\alpha }%
_{i}\leftrightarrow \alpha _{i}$ . (\ref{Postalin}) is a remarkable
equation: Since the mass parameters $m_{i}$ in the $\mathcal{N}=2$ gauge
theory are given by $\alpha _{i}=\frac{Q}{2}+m_{i}$, from (\ref{a2})-(\ref%
{a4}) and (\ref{Postalin}) we see how the symmetry under transformation (\ref%
{c1})-(\ref{c4})\ is realized in the Liouville theory picture (cf. Eqs.
(4.1)-(4.3) of Ref. \cite{AGT}). Indeed, defining 
\begin{eqnarray}
X(\alpha _{1},\alpha _{2},\alpha _{3},\alpha _{4}|q) &=&|q|^{4\alpha
_{1}\alpha _{4}}|1-q|^{4\alpha _{2}\alpha _{4}}\left( \pi \mu \gamma
(b^{2})b^{2-2b^{2}}\right) ^{\frac{\alpha _{1}+\alpha _{2}+\alpha
_{3}-\alpha _{4}}{2b}}\times   \notag \\
&&\prod_{i=1}^{4}\Upsilon _{b}^{-1}(2\alpha _{i})\ \ \Omega ^{(4)}(\alpha
_{1},\alpha _{2},\alpha _{3},\alpha _{4}|0,1,\infty ,q).
\end{eqnarray}%
we have 
\begin{equation}
X(\alpha _{1},\alpha _{2},\alpha _{3},\alpha _{4}|q)=X(\widetilde{\alpha }%
_{1},\widetilde{\alpha }_{2},\widetilde{\alpha }_{3},\widetilde{\alpha }%
_{4}|q).  \label{sdw}
\end{equation}

Identity (\ref{Postalin}) was \textit{a priori} unexpected, and the fact
this codifies precisely the symmetry under (\ref{c1})-(\ref{c4}) in the $%
\mathcal{N}=2$ gauge theory side is interesting as it permits to understand
the SO($8$) symmetry of the theory from the Liouville theory point of view
(and not only the SO($4$)$\times $SO($4$) proper subgroup.)

Now, let us make some comments on surface operators in the gauge theory. We
will come back to Eq. (\ref{Postalin}) later.

\section{Surface operator and WZW\ theory}

\subsection{Surface operators from WZW correlators}

It was argued in \cite{AGT2,Teschner} that the expectation value of loop and
surface operators in the gauge theory can also be described in terms of
Liouville correlation functions. While loop operators are described by a
monodromy operation performed on a degenerate field $V_{-1/2b}=e^{-\varphi /(%
\sqrt{2}b)},$ the expectation values of elementary surface operators are
given in terms of the correlation function%
\begin{equation*}
\Omega ^{(n+1)}(\alpha _{1},...\alpha _{n},-1/(2b)|z_{1},...z_{n},x)\
=\left\langle \prod\nolimits_{i=1}^{n}V_{\alpha _{i}}(z_{i})\ V_{-\frac{1}{2b%
}}(x)\right\rangle _{\text{Liouville}}
\end{equation*}%
on a ($n+1$)-puncture genus-$g$ Riemann surface, which also involves a
degenerate field $V_{-1/2b}$ that is inserted at the point $x$, and $x$ is
related to the parameters that label the corresponding gauge theory
configuration.

These surface operators correspond to 1/2 BPS configurations in the $%
\mathcal{N}=2$ theory. The analogue in the $\mathcal{N}=4$ theory is a
singular vortex solution, which is labeled by two real parameters that
correspond to its magnetic flux and a $\theta $-angle type parameter.
Typically, there are always two real parameters that correspond to physical
quantities and label the corresponding vortex-like configuration in the
gauge theory, and in the Liouville picture these two real parameters combine
to give the complex worldsheet coordinate $x$ where the degenerate operator $%
V_{-1/2b}(x)=e^{-\varphi (x)/(\sqrt{2}b)}$ is inserted. Surface operators
correspond to configurations that are localized on the two-dimensional
Riemann surface $\mathcal{C}_{g,n}$; this is discussed in detail in \cite%
{AGT2,Teschner,GaiottoNuevo}.

The degenerate state created by the non-normalizable vertex operator $%
V_{-1/2b}$ is annihilated by the arrange of Virasoro operators 
\begin{equation}
\left( L_{-1}^{2}+b^{2}L_{-2}\right) V_{-\frac{1}{2b}}=0.  \label{Q}
\end{equation}%
This expresses the fact that a null state exists in the Verma modulo and it
has to be decoupled. Actually, (\ref{Q}) can be thought of as a realization
of the second order equation of motion%
\begin{equation}
\left( \partial _{z}^{2}+T(z)\right) e^{-\varphi /2}=0  \label{C}
\end{equation}%
at quantum mechanical level. Decoupling equation (\ref{Q}) has to be
understood as an operator-valued relation which, when implemented on
correlation functions, yields a second order differential equations of the
Belavin-Polyakov-Zamolodchikov (BPZ) type \cite{BPZ} to be obeyed by those
correlators that involve a degenerate field with $\alpha _{n+1}=-1/2b$.

On the other hand, one knows from \cite{FZ} that given a solution $\Omega
^{(5)}(\alpha _{1},...\alpha _{4},-1/2b|0,1,\infty ,z,x)$ to the five-point
BPZ differential equation one can associate to it a solution $%
K(j_{1},...j_{4}|q,x)$ to the four-point Knizhnik-Zamolodchikov (KZ)
differential equation \cite{KZ} with affine symmetry $\widehat{A}_{1}$ at
level $k=b^{-2}+2$. The exact relation is given by%
\begin{equation}
K(j_{1},...j_{4}|q,x)=\mathcal{N}\ \frac{|q|^{4(\alpha _{1}\alpha
_{4}-b^{2}j_{1}j_{4})}|q-1|^{4(\alpha _{2}\alpha _{4}-b^{2}j_{2}j_{4})}}{%
|x|^{2\alpha _{1}/b}|x-1|^{2\alpha _{2}/b}|x-q|^{2\alpha _{4}/b}}\ \Omega
^{(5)}(\alpha _{1},...\alpha _{4},-1/2b|0,1,\infty ,q,x)  \label{karita}
\end{equation}%
where 
\begin{eqnarray}
\alpha _{1} &=&-\frac{b}{2}(j_{1}+j_{2}+j_{2}+j_{4}+1),\qquad \alpha _{2}=-%
\frac{b}{2}(-j_{1}+j_{2}-j_{3}+j_{4}-b^{-2}-1),  \label{R1} \\
\alpha _{3} &=&-\frac{b}{2}(-j_{1}-j_{2}+j_{3}+j_{4}-b^{-2}-1),\qquad \alpha
_{4}=-\frac{b}{2}(j_{1}-j_{2}-j_{3}+j_{4}-b^{-2}-1),  \label{R2}
\end{eqnarray}%
\ and where $\mathcal{N}$ is a normalization factor that does not depend on (%
$x$,$q$), and where $\alpha _{5}=-1/(2b)$, $b^{-2}=k-2$, $z_{1}=0$, $z_{2}=1$%
, $z_{3}=\infty $, and $z_{4}=q$. Notice that (\ref{R1})-(\ref{R2})
resembles (\ref{a2})-(\ref{a4}) and (\ref{c1})-(\ref{c4}); we will make this
more precise; see (\ref{HR}) below.

If the normalization $\mathcal{N}$ is taken to be \cite{T} 
\begin{equation}
\mathcal{N}=\mu ^{2j_{4}}\left( \pi \gamma (b^{2})b^{2-2b^{2}}\right)
^{2j_{4}+2\alpha _{4}/b}\prod_{i=1}^{4}\frac{\Upsilon _{b}(-2bj_{i}-b)}{%
\Upsilon _{b}(2\alpha _{i})}  \label{NNNNNNNNNNNNN}
\end{equation}%
then the left hand side of (\ref{karita}) can be interpreted as the
four-point correlation function of the level-$k$ $\widehat{A}_{1}$ WZW
theory; namely

\begin{equation}
K(j_{1},...j_{4},|q,x)=\left\langle \prod\nolimits_{i=1}^{4}\Phi
_{j_{i}}(x_{i}|z_{i})\right\rangle _{\text{WZW}},  \label{YESTA}
\end{equation}%
up to an irrelevant $b$-dependent factor.

It is worth mentioning that an expression similar to (\ref{karita}) holds at
the level of conformal blocks \cite{T}, and not only for the full
correlation function. Although (\ref{karita}) is usually interpreted as a
relation between Liouville theory and $\widehat{sl}(2)_{k}$ WZW theory, it
is pertinent to emphasize that the $\widehat{su}(2)_{k}$ WZW appears by
analytic extending the expressions reversing the sign of $k$; see \cite%
{Lorena} and references therein.

Vertex operators $\Phi _{j}(x|z)$ in (\ref{YESTA}) represent states of the
WZW\ theory. These are given by Kac-Moody primary states with respect to the
affine $\widehat{A}_{1}$ symmetry. These vertices expand $SL(2,\mathbb{R})$%
-representations of spin $j_{i}$, and depend on auxiliary complex variables $%
x_{i}$ which allow to organize the representations by means of the following
realization%
\begin{equation*}
J^{a}(z)\Phi _{j}(x|w)=-\frac{\mathcal{D}_{x}^{a}\Phi _{j}(x|w)}{(z-w)}+...
\end{equation*}%
with differential operators%
\begin{equation*}
\mathcal{D}_{x}^{+}=x^{2}\partial _{x}-2jx,\quad \mathcal{D}%
_{x}^{-}=\partial _{x},\quad \mathcal{D}_{x}^{3}=x\partial _{x}-j,
\end{equation*}%
where, as usual, the notation $a=+,-,3$ refers to the indices of the
currents $J^{\pm }(z)=J^{1}(z)\pm iJ^{2}(z)$ and $J^{3}(z)$, which generate
the affine $\widehat{sl}(2)_{k}$ algebra.

According to (\ref{karita})-(\ref{YESTA}), the expectation value of
elementary surface operators in the $\mathcal{N}=2$ SCFT with $N_{f}=4$ is
given by a WZW\ four-point function. In some sense, the presence of a theory
with affine $\widehat{A}_{1}$ symmetry seems to be natural from the $%
\mathcal{N}=2$ theory point of view. In fact, one expects the Riemann
surface to have SU($2$) structure at the punctures. This suggests that
trying to see AGT correspondence from the WZW perspective could be useful to
understand the connection between 4D and 2D theories in more detail. We will
comment on the WZW model at the end of this section and we will suggest that
it could play an important r\^{o}le in this story.

\subsection{Alternative derivation of (\protect\ref{Postalin})}

What we would like to discuss now is another consequence of relation (\ref%
{Postalin}). We will see how (\ref{Postalin}), together with (\ref{karita}),
enable to find a nice relation between correlation functions of Liouville
theory and of WZW\ theory. To see this, first consider the special operator
product expansion 
\begin{equation}
V_{\alpha _{i}}(z_{i})V_{-1/2b}(x)\underset{z_{i}\rightarrow x}{=}C_{-}\
\left| x-z_{i}\right| ^{2\xi _{-}}V_{-1/2b+\alpha _{i}}(z_{i})+C_{+}\ \left|
x-z_{i}\right| ^{2\xi _{+}}V_{-1/2b-\alpha _{i}}(z_{i}),  \label{ma}
\end{equation}%
with%
\begin{equation*}
C_{+}=(\pi \mu \gamma (b^{2}))^{b^{-2}}\frac{\gamma (2\alpha
_{i}b^{-1}-1-b^{-2})}{b^{4}\gamma (2\alpha _{i}b^{-1})},\qquad C_{-}=1,
\end{equation*}%
where $\xi _{\pm }=(\Delta _{\alpha _{i}\pm 1/2b}-\Delta _{1/2b}-\Delta
_{\alpha _{i}})$. OPE (\ref{ma}) is an important piece of information about
the structure of Liouville theory \cite{Teschner3pf,Pakman}: It permits to
write the coincidence limit of the Liouville five-point function $\Omega
^{(5)}(\alpha _{1},\alpha _{2},\alpha _{3},\alpha _{4},-1/(2b)|0,1,\infty
,q,x)$ in terms of only two (i.e. not infinite) four-point contributions $%
\left| x-q\right| ^{2\xi _{\mp }}\Omega ^{(4)}(\alpha _{1},\alpha
_{2},\alpha _{3},-1/(2b)\pm \alpha _{4}|0,1,\infty ,q)$. And here is where
equation (\ref{Postalin}) comes to play an important r\^{o}le: Using the
fact that $\Omega ^{(4)}(\alpha _{1},\alpha _{2},\alpha _{3},\alpha
_{4}|0,1,\infty ,q)$ and $\Omega ^{(4)}(\widetilde{\alpha }_{1},\widetilde{%
\alpha }_{2},\widetilde{\alpha }_{3},\widetilde{\alpha }_{4}|0,1,\infty ,q)$
are connected in such a direct way, one rapidly finds the remarkable relation%
\begin{equation}
K(j_{1},...j_{4}|q,x)\underset{x\rightarrow q}{\simeq }\prod%
\nolimits_{i=1}^{4}\frac{1}{\gamma (-b^{2}(2j_{i}+1))}\ \Omega
^{(4)}(-bj_{1},-bj_{2},-bj_{3},-bj_{4}|0,1,\infty ,q),  \label{HR}
\end{equation}%
which, in contrast to (\ref{karita}), connects Liouville four-point
correlation functions to WZW four-point correlation functions. Symbol $%
\simeq $ stands here because a regularization is needed due to a singular
factor $\left| x-q\right| ^{-2(1+b^{-2})}$ that blows up in the limit $%
x\rightarrow q$. Subleading contribution vanish in the limit, provided the
Seiberg bound $\alpha _{i}>Q/2$ is obeyed \cite{Seiberg}. Notice that
factors $\gamma (1+b^{2}(2j_{i}+1))$ in (\ref{HR}) can be absorbed in the
normalization of WZW vertices, in such a way that (\ref{HR}) induces a
natural one-to-one identification between fields $\Phi
_{j}(z_{i})\leftrightarrow V_{-bj}(z_{i})$ of both theories, cf. (\ref%
{karita}). The fact that (\ref{HR}) takes such a simple factorized form is
due to remarkable cancellations that occur through the calculation.

Expression (\ref{HR})\ provides us with a nice realization of the so-called
Drinfeld-Sokolov Hamiltonian reduction WZW $\rightarrow $\ Liouville$\ $at
the level of the four-point correlation functions. From the gauge theory
point of view, and according to the interpretation of the degenerate field $%
V_{-1/2b}$ of Liouville theory as representing the surface operator of the
gauge theory \cite{AGT2,Teschner}, expression (\ref{HR}) means that the
limit $x\rightarrow q$ of the expectation value of a surface operator in the 
$\mathcal{N}=2$ theory is indeed described by a WZW correlation function.
Understanding the physical picture behind this fact requires further
investigation. In particular, understanding the relation of (\ref{HR})
within the context of the Langlands correspondence \cite{Frenkel} is matter
of further work.

Another interesting aspect about (\ref{HR}) is that, even though here we
used (\ref{Postalin}) to give a concise proof of it, it is believed to hold
independently as it follows from the Hamiltonian reduction realized at the
level of correlation functions; see for instance \cite{Rasmussen}. This
means that, without risk of circular arguments, we can consider (\ref{HR})
as the starting point and then use it to prove identity (\ref{Postalin}),
reversing the story. In fact, once one assumes (\ref{HR}), Eq. (\ref%
{Postalin}) simply follows from functional relations between different
solutions of the KZ\ equation that were proven in Refs. \cite%
{Giribet2005,Giribet2006}. It is easy to see that some of the relations
discussed in \cite{Giribet2005}, together with Weyl symmetry, lead to
identify solutions $K(j_{1},...j_{4}|q,x)$ with solutions $K(\widetilde{j}%
_{1},...\widetilde{j}_{4}|q,x)$, where $\widetilde{j}_{i}$ are related to $%
j_{i}$ through (\ref{a2})-(\ref{a4}) recalling $j_{i}=-\alpha _{i}/b.$
Nevertheless, the way we proved equation (\ref{Postalin}) in Section 2 is
more direct and, consequently, is the one we prefer.

\subsection{More comments on WZW theory}

Before concluding, and since we are already talking about the relation
between Liouville and WZW theories, let us make some comments on another
connection that exists between these two conformal theories, and which
probably may have important implications for AGT.

It was shown in \cite{Stoyanovsky,RibaultTeschner} that $n$-point $\widehat{%
sl}(2)_{k}$ WZW correlation functions on the sphere are equivalent to ($2n-2$%
)-point correlation functions in Liouville theory, where $n-2$ states in the
Liouville correlators are degenerate fields $V_{-1/2b}$. In \cite%
{HikidaSchomerus} the result of \cite{RibaultTeschner} was generalized to
genus-$g$ correlation functions. According to the results of \cite%
{HikidaSchomerus}, any $n$-point function in WZW theory on a genus-$g$
surface is equivalent to a ($2n+2g-2$)-point function in Liouville theory
which involves $n+2g-2$ degenerate fields. This means that Liouville
correlation function on a surface $\mathcal{C}_{g,n}$ with one additional
field $V_{-1/2b}$ for each trinion that appears in a given way of sewing the
surface, is actually equivalent to a $n$-point WZW correlator. Whether this
observation has some deep meaning from the gauge theory point of view is an
open question.

The correspondence between Liouville and WZW theories found in \cite%
{Stoyanovsky,RibaultTeschner}\ was further extended in \cite{Ribault} to the
case of $SL(2,\mathbb{R})$ primary states of spectral flowed sectors; this
yields a correspondence between $n$-point WZW\ correlators and $m$-point
Liouville correlators, where $n-m$ is the total amount of spectral flow
units in the WZW observable. A similar relation was proposed between
correlators of Liouville theory that involve higher-level degenerate fields $%
V_{-m/2b}$ and correlators in a yet-to-be explored family of non-rational
conformal field theories with central charges given by $%
c_{(m)}=3+6(b+b^{-1}(1-m))^{2}$, \cite{RibaultFamily}. This could lead to
describe higher-monodromy loop operators of the $\mathcal{N}=2$ theories in
terms of such new family of CFTs. Furthermore, it is a common belief that a
generalization of the WZW-Liouville correspondence exists between the $%
\widehat{sl}(N)_{k}$ WZW theory and higher-rank Toda field theory (see \cite%
{RibaultSL3} for recent attempts in this direction). Speculatively, this
could lead to a description of SU($N_{c}$) $\mathcal{N}=2$ theories in terms
of the WZW with affine symmetry $\widehat{A}_{N_{c}-1}$. The results of \cite%
{Stoyanovsky,RibaultTeschner,HikidaSchomerus} permit to describe observables
in the $\mathcal{N}=2$ gauge theories in terms of WZW correlators. However,
the physical meaning of this still remains elusive. Likely, this picture
will eventually permit to extend AGT correspondence and establish a more
general connection between gauge theories and two-dimensional conformal
theories. In \cite{HikidaSchomerus} the connection between the Liouville-WZW
correspondence and the Langlands correspondence was pointed out; see also 
\cite{YuYo}. Investigating this from the gauge theory perspective \cite%
{KapustinWitten} is an interesting project for future investigations.

\begin{equation*}
\end{equation*}

This work was partially supported by University of Buenos Aires, Agencia
ANPCyT, and CONICET, through grants UBACyT X861, UBACyT X432,
PICT-2007-00849. The author thanks L.F. Alday and D. Gaiotto for helpful
discussions, and he specially thanks Y. Tachikawa for pointing out an
important mistake in the first version of the preprint. The author also
thanks J. Edelstein, Y. Hikida, Yu Nakayama and J. Teschner for very
interesting comments, and thanks V. Fateev for pointing out reference \cite%
{FL2}. The author is grateful to M. Kleban, M. Porrati, and the members of
the Center for Cosmology and Particle Physics of New York University NYU for
their hospitality.

\end{document}